\def\beq{\begin{equation}}
\def\eeq{\end{equation}}
\def\beqa{\begin{eqnarray}}
\def\eeqa{\end{eqnarray}}
\def\bes{\begin{split}}
\def\ees{\end{split}}
\def\ua{\uparrow}
\def\da{\downarrow}
\newcommand{\bi}{Bi$_2$Ba$_2$CaCu$_2$O$_y$}
\newcommand{\cuo}{CuO$_2$ }

\newcommand{\flrin}{Ba$_2$Ca$_3$Cu$_4$O$_8$(O$_{1-y}$F$_y$)$_2$}
\newcommand{\hg}{HgBa$_2$Ca$_4$Cu$_5$O$_y$}

\newcommand{\para}{\parallel}

\newcommand{\tj}{{\it t-t'-t''-J} }

\newcommand{\tc} {$T_{\rm c}$}
\newcommand{\tpp}{$t_{\perp}$}

\def\jnl#1#2#3#4{#1 {\bf #2} (#4) #3.}

\def\EPJB{Eur.\ Phys.\ J. B}
\def\IJMPB{Int.\ J.\ Mod.\ Phys. B}

\def\JPSJ{J.\ Phys.\ Soc.\ Jpn.}

\def\PR{Phys.\ Rev.}
\def\PRB{Phys.\ Rev.\ B}
\def\PRL{Phys.\ Rev.\ Lett.}
\def\PTP{Prog.\ Theor.\ Phys.}
\def\RMP{Rev.\ Mod.\ Phys.}

\def\ZPB{Z.\ Phys.\ B}
\documentclass[twocolumn]{jpsj2} %% two-column layout
\title{Charge Imbalance Effects on Interlayer Hopping and Fermi Surfaces in Multilayered High-$T_{\rm c}$ Cuprates}

\author{M. Mori$^{1}$, T. Tohyama$^{1}$ and S. Maekawa$^{1,2}$}

\inst{	$^{1}$Institute for Materials Research, Tohoku University, Sendai 980-8577, Japan\\
	$^{2}$CREST, Japan Science and Technology Agency, Kawaguchi 433-0012, Japan}

\abst{
We study doping dependence of interlayer hoppings, $t_{\perp}$, in multilayered cuprates with four or more \cuo planes in a unit cell. 
When the double occupancy is forbidden in the plane, an effective amplitude of $t_{\perp}$  in the Gutzwiller approximation is shown to be proportional to the square root of the product of doping rates in adjacent two planes, i.e., 
$t^{\rm eff}_{\perp}\propto t_{\perp}\sqrt{\delta_{\rm 1}\delta_{\rm 2}}$, where $\delta_1$ and $\delta_2$ represent the doping rates of the two planes. 
More than three-layered cuprates have two kinds of \cuo planes, i.e., inner- and outer planes (IP and OP), resulting in two different values of $t^{\rm eff}_{\perp}$, i.e., 
$t^{\rm eff}_{\perp1}\propto t_{\perp}\sqrt{\delta_{\rm IP}\delta_{\rm IP}}$ 
between IP's, and 
$t^{\rm eff}_{\perp2}\propto t_{\perp}\sqrt{\delta_{\rm IP}\delta_{\rm OP}}$ 
between IP and OP. 
Fermi surfaces are calculated in the four-layered $t$-$t'$-$t''$-$J$ model by the mean-field theory. The order parameters, the renormalization factor of $t_{\perp}$, and the site-potential making the charge imbalance between IP and OP are self-consistently determined for several doping rates. 
We show the interlayer splitting of the Fermi surfaces, which may be observed in the angle resolved photoemission spectroscopy measurement. 

}
\kword{multilayered cuprates, interlayer coupling, Gutzwiller approximation, t-J model, ARPES}

\begin{document}
\maketitle

\section{Introduction}
Cuprate superconductors have layered structure of \cuo planes, which constitute conducting blocks separated by charge-reservoir blocks. 
Multilayered high-\tc cuprates, e.g., \flrin\cite{iyo01,iyo03,iyo04} and \hg\cite{schwer,capponi,akimoto}, have two kinds of \cuo planes in one conducting block; the pyramidally-coordinated-outer-planes (OP's) and the square-coordinated-inner-planes (IP's). 
Due to spatial variation of Madelung potential\cite{kondo,distatio,ohta,haines}, the carrier density in the OP is generally different from that in the IP. 
We call such an inhomogeneous charge-distribution $\lq$charge imbalance'. 

The charge imbalance has been observed by the nuclear magnetic resonance (NMR) measurement\cite{trokiner,statt,magishi,zheng,julien,tokunaga,kotegawa01} and the X-ray diffraction measurement with the bond-valence analysis\cite{karppinen1,karppinen2,fujii}. 
As regards hole-doped systems, the NMR studies have found that a hole concentration in the OP, $\delta_{\rm OP}$, is larger than that in the IP, $\delta_{\rm IP}$. 
A difference of doping rate, $\Delta \delta =\delta_{\rm OP}-\delta_{\rm IP}$, increases with the number of \cuo planes in one conducting block, $n$\cite{kotegawa01}. 
Although the five-layered cuprates have three IP's, the charge imbalance among them is small compared to $\Delta \delta$. 

It is known that the superconducting critical temperature, \tc, increases with $n$ and saturate at $n=3$\cite{karppinen1,karppinen2}. 
Thereafter, \tc~decreases with $n$, while the charge imbalance increases. 
Considered such an $n$-dependence of \tc, the charge imbalance seems to suppress \tc~for $n>$3. 
Simultaneously, the charge imbalance induces some interesting phases, e.g., two kinds of superconducting (SC) gaps\cite{tokunaga} and coexistence of the SC- and antiferromagnetic (AF) states\cite{kotegawa04,tokiwa}. 
In the former case, the OP's have a large SC gap compared to the IP's. 
In the latter case, the SC OP's are separated by the AF IP's, and a long-ranged Josephson coupling through the AF planes stabilizes the SC state as a bulk\cite{mori_lett}. 
The suppression of \tc~may be partly caused by these inhomogeneous phases.

On the other hand, some theoretical studies have been presented that an interlayer hopping plays an important role about the $n$-dependence of \tc\cite{kivelson,chak_nat,zaleski,mori_lett}. 
The interlayer hopping is indispensable for the bulk SC state, even if each \cuo plane shows superconductivity\cite{anderson,tesanovic,wheatley,chak_sci,leggett,chak_epjb,emery}. 
Here, we note two kinds of interlayer hoppings in the multilayered cuprates. 
One is a coherent hopping between adjacent planes within the conducting block, and the other one is an incoherent hopping through the charge reservoir block. 
The 'coherent' means the conservation of momentum. 
Although both contribute to stabilize the bulk superconductivity, it is the former interlayer hopping that gives rise to the $n$-dependence of \tc. 
Below, we will focus our attention on the coherent hopping within the conducting block and simply call it 'interlayer hopping'. 
 
An amplitude of the interlayer hopping, \tpp, has been observed in the bilayer cuprates, e.g., \bi, by the angle-resolved-photoemission-spectroscopy (ARPES) measurement\cite{feng,chuang}.
The experimental estimation of \tpp~in the bilayer one is about 50 meV\cite{feng}, which is much smaller than 150 meV  obtained by the bilayer LDA calculations\cite{okandersen}. On the other hand, the bilayer Hubbard model predicted a similar amplitude of 40 meV\cite{liechtenstein}. 
This is because a strong on-site Coulomb repulsion in the \cuo plane substantially reduces \tpp\cite{anderson_prl,eder}. 
Furthermore, the interlayer hopping depends on the doping rate as well. For example, in the bilayer case, the splitting of Fermi surface (FS) is clear in the overdoped region, while it is not observed in the underdoped region.\cite{feng,chuang}
Therefore, as regards the multilayered cuprates for $n\ge$4, the relation between the interlayer hoppings and the Fermi surface splittings are not so obvious. 
One can suppose that the charge imbalance gives spatially inhomogeneous \tpp's, i.e., \tpp~between IP's is different from that between IP and OP.
Such inhomogeneous interlayer hoppings lead to unexpected FS splittings, which will tell another key factor of $n$-dependence of \tc~for $n>$ 3. 

In this paper, we study doping dependence of $t_{\perp}$ to show the FS splitting in the multilayered cuprates. 
The effective \tpp~renormalized by the on-site Coulomb repulsion is calculated in the Gutzwiller approximation (GWA)\cite{gutzwiller63,gutzwiller64,gutzwiller65,ogawa75,ogawa78,vollhardt,rice_prl,rice_prb,yokoyama87,yokoyama96,zhang,hsu,sigrist,ogata}. 
It has been shown that the GWA gives a reliable estimate of the variational energy for the metallic- and SC states\cite{yokoyama87,yokoyama96,zhang}, while extensions of the GWA are necessary for the AF state\cite{hsu,sigrist,ogata}. 
Below, we consider the metallic \cuo planes without the AF order. 
When the double occupancy is forbidden in each plane, we have found that the renormalization factor, $q_{\perp}$, is proportional to the square root of the product of the doping rates in two planes, resulting in two different values of $q_{\perp}$, i.e., 
$q_{\perp,1}\propto \sqrt{\delta_{\rm IP}\delta_{\rm IP}}$ 
between IP's, and 
$q_{\perp,2}\propto \sqrt{\delta_{\rm IP}\delta_{\rm OP}}$ 
between IP and OP. 
Taking account of $q_{\perp,1}$ and $q_{\perp,2}$, FS's are calculated in the multilayered $t$-$t'$-$t''$-$J$ model in the mean-field theory. 
The order parameters, the renormalization factors, and the site-potential making the charge imbalance are self-consistently determined for several doping rates. 
We find that the FS splittings are not always well separated, e.g., only two FS in the four-layered system. 

The rest of this paper is organized as follows.
In \S2, we briefly summarize the Gutzwiller approximation for the interlayer hopping between two planes with different doping rates.
In \S3, the FS's are calculated for the four-layered case of the multi-layered \tj model including the inter-layer hopping obtained in the previous section.  
Summary and discussions are given in \S4.

\section{Gutzwiller Factor of  Interlayer Hopping}
We consider two planes with the charge imbalance given by 
\beqa
H 
	&=& 
	H_{\parallel} + H_W + H_{\perp},\\
H_{\parallel}
	&=&
	\sum_{l=1,2}\left(t^{\para}\sum_{i,j; \sigma}c_{i\sigma}^{(l)\dag}c^{(l)}_{j\sigma}
   + U\sum_i n^{(l)}_{i\ua} n^{(l)}_{i\da}\right),\\
H_W
	&=&
		\frac{W}{2}\sum_i \left(n^{(1)}_i - n^{(2)}_i \right),\label{sitepot}\\
H_{\perp} 
	&=&
  t^{\perp} \sum_{i,j} 
	\left( c_{i\sigma}^{(1)\dag}c^{(2)}_{j\sigma}+H.c.\right),
\eeqa
where $c^{(l)\dag}_{i\sigma}$ ($c^{(l)}_{i\sigma}$) is the electron creation (annihilation) operator with spin $\sigma$ at the $i$-th site in the $l$-th plane, and  
$n^{(l)}_{i\sigma} = c_{i\sigma}^{(l)\dag} c^{(l)}_{i\sigma}$ and  
$n^{(l)}_i = n^{(l)}_{i\ua} + n^{(l)}_{i\da}$. 
Hopping integral in the $l$-th plane and a strength of Coulomb repulsion are denoted by $t^{\para}$ and $U$, respectively. 
Several cases of the charge imbalance are schematically shown in Fig. 1. The hole density is denoted by $\delta$, and the density of double occupancy is indicated by $d$. 
%******
\begin{figure}[p]
\begin{center}
\includegraphics[width=6.5cm]{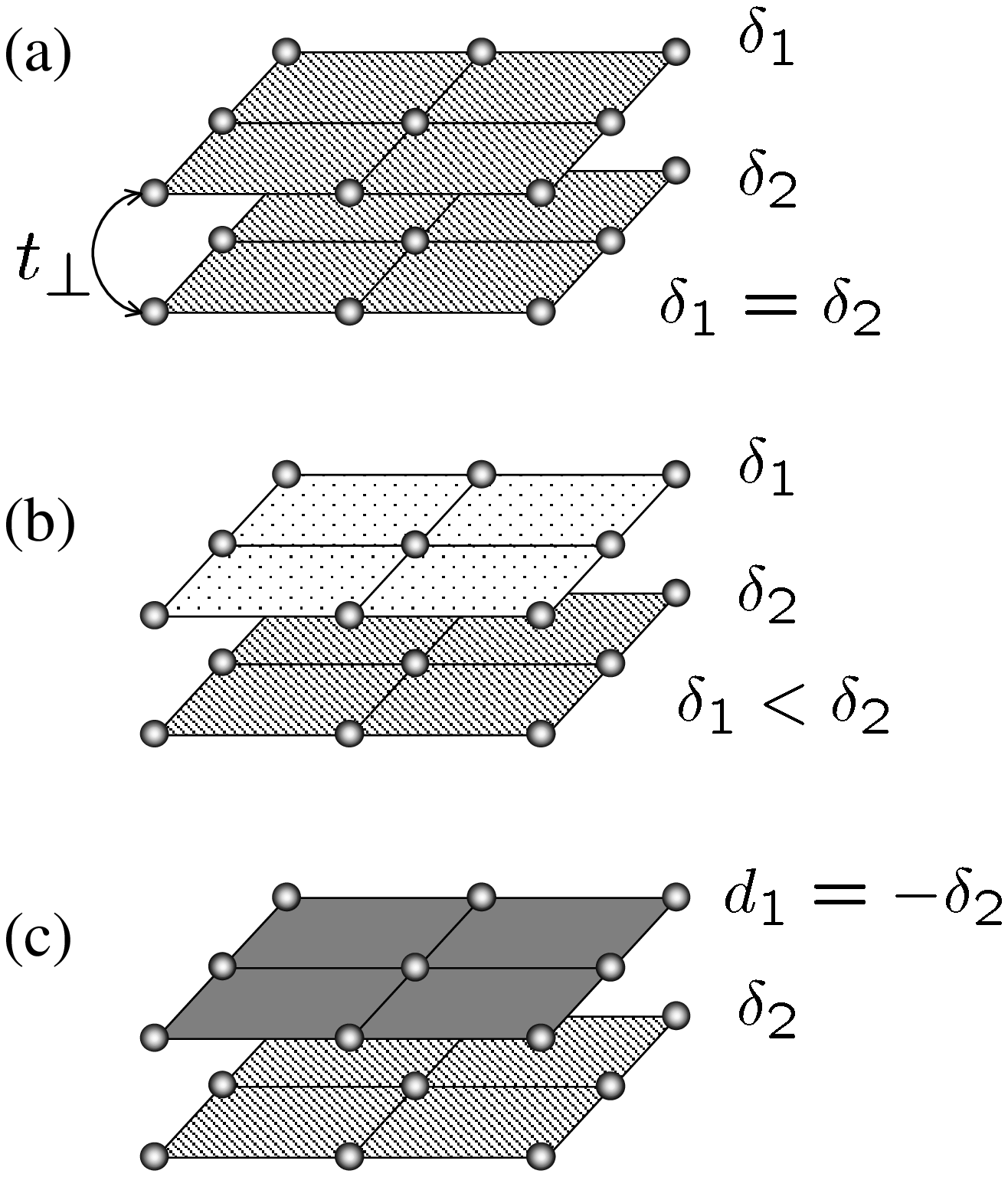}
%\vspace{10cm}
\caption{(a) equivalent two planes with hole-doping, (b) hole-doped two planes with charge imbalance, (c) the one is hole-doped and the other one is electron-doped. }
\label{fig1}
\end{center}
\end{figure}
%******

%---------------------------------- 
%\section{Gutzwiller wavefunction}
%---------------------------------- 
The Gutzwiller's wavefunction is assumed for the ground state as
\beq
|\psi\rangle =
	 P_n P_g|\psi_0\rangle,
\eeq
where $P_n$ is the projection operator onto the subspace of fixed densities, $n^{(1)}$ and $n^{(2)}$. 
The Gutzwiller's projection operator, $P_g$, with a variational parameter, $g_l$, is given by 
\beq
P_g =
	\prod_{i;l=1,2} \left\{1-(1-g_{l})n^{(l)}_{i,\ua}n^{(l)}_{i,\da}\right\}, 
\eeq
and
$|\psi_0\rangle$ is a ground state wavefunction for $U=0$. 

%---------------------------------- 
%\section{Effective Hamiltonian and Ground state energy}
%----------------------------------
The ground state energy, $E_g$, is evaluated as  
\beq
E_{\rm g} 
	=
	\frac{\langle \psi |H|\psi\rangle}
		    {\langle \psi | \psi\rangle}. 
\eeq
In the GWA, $P_g$ is replaced by classical weighting factors, which is called the Gutzwiller factor. 
Accordingly, $E_g$ is calculated by the expectation value of an effective Hamiltonian as,   
\beqa
&&E_{\rm g} 
	=
	\langle \psi_0 |H_{\rm eff} |\psi_0 \rangle + U(d_1+d_2),\\
&&H_{\rm eff}(n^{(1)},n^{(2)})	
	= 
	\sum_{l=1,2}\left(q_t^{(l)} t^{\para}\sum_{i,j,\sigma} c_{i\sigma}^{(l)\dag}c^{(l)}_{j\sigma}\right)\nonumber\\
	&&+
		\frac{W}{2}\sum_i \left(n^{(1)}_i - n^{(2)}_i \right)
	+
   q_{\perp}t^{\perp}\sum_{i,j,\sigma} 
	\left(c^{(1)\dag}_{i\sigma}c^{(2)}_{j\sigma}+H.c.\right),\\
&&d_{l}
	\equiv
		\langle n^{(l)}_{i\ua}n^{(l)}_{i\da}\rangle,
\eeqa
where $d_{l}$ is a density of doubly occupied sites in the $l$-th plane.
The Gutzwiller factors of the intraplane- and the interlayer hoppings are denoted by $q_t^{(l)}$ and $q_{\perp}$, respectively. 

%----------------------------------
%\subsection{Intersite correlation effect}
%----------------------------------
In the original GWA, only the site-diagonal expectation values, $\langle n_{i\sigma}\rangle$, are considered. 
In contrast, the intersite correlation, $\langle c^{\dag}_{i\sigma}c_{j\sigma}\rangle$, becomes important to obtain $q_t^{(l)}$ in the AF state\cite{ogawa78,hsu,sigrist,ogata} having the two sublattices. 
Then, it is interesting whether the intersite correlation for $q_{\perp}$ is important in the paramagnetic state with charge imbalance. 
In this case, however, the intersite correlation is negligible for $q_{\perp}$. 
Details are summarized in the Appendix. 

%---------------------------------- 
%\section{Gutzwiller factor}
%----------------------------------
In the way of original GWA, $q_t^{(l)}$ is given by\cite{vollhardt}
\beqa
&&q_t^{(l)} =
	\frac{
		\left(
			\sqrt{(n_{l}-d_{l})(1-2n_{l}+d_{l})}+\sqrt{d_{l}(n_{l}-d_{l})}
		\right)^2}
		{n_{l}(1-n_{l})}, \\
&&n_{l} \equiv 
	\langle n^{(l)}_{i\ua}\rangle = \langle n^{(l)}_{i\da}\rangle.
\eeqa
Even in the site-diagonal approximation, $q_{\perp}$ is not so obvious, since the two planes have different carrier densities\cite{rice_prl,rice_prb}. 
The renormalization of the interlayer hopping from the $l$-th plane to the $m$-th one is different from that of the conjugate process as,
\beqa
\frac{\langle \psi |c^{(m)\dag}_{i\sigma}c^{(l)}_{j\sigma}|\psi\rangle}
	    {\langle \psi | \psi\rangle}
	&=&
		q_{l\to m}\langle \psi_0 |c^{(m)\dag}_{i\sigma}c^{(l)}_{j\sigma}|\psi_0\rangle,
\eeqa
where
\beqa
q_{l\to m}
	&=&
	\frac{(n_l - d_m)}{n_l(1-n_m)}
	\Bigl\{
	(1-2n_m+d_m) \nonumber\\
&&\left.+ \sqrt{d_m(1-2n_m+d_m)}\right.\nonumber\\
&&\left.+ (1-2n_m+d_m)\sqrt{\frac{d_l}{(1-2n_l+d_l)}}\right.\nonumber\\ 
&&\left.+ \sqrt{d_l c_m\frac{(1-2n_m+d_m)}{(1-2n_l+d_l)}}
	\right\}.
\eeqa
The GW factor for the interlayer hopping can be obtained as,
\beqa
q_{\perp}
	&=&
	\sqrt{q_{1\to 2}q_{2\to 1}} = \sqrt{q_1q_2},\\
	&=&
		\left\{\frac{	(n_1-d_1)(n_2-d_2)}{n_1(1-n_1)n_2(1-n_2)}\right\}^{1/2}\nonumber\\
	&&\times
		\Bigl[
			\{d_1+(1-2n_1+d_1)\}\{d_2+(1-2n_2+d_2)\}\nonumber\\
			&&\left.
			+ 2\{d_1+(1-2n_1+d_1)\}\sqrt{d_2(1-2n_2+d_2)}\right.\nonumber\\
			&&\left.
			+ 2\{d_2+(1-2n_2+d_2)\}\sqrt{d_1(1-2n_1+d_1)}\right.\nonumber\\
			&&\left. 
			+4\sqrt{d_1d_2(1-2n_1+d_1)(1-2n_2+d_2)}
		\right]^{1/2}.\label{qperp}
\eeqa

%---------------------------------- 
%\section{case studies of interlayer GW factor}
%----------------------------------
%\subsection{Hole- and hole-doped layers}
%----------------------------------
If one considers two hole-doped planes without double occupancy, i.e., $d_1=d_2=0$, the interlayer GW factor is given by
\beqa
q_{\perp}
	&=&
		\left(\frac{	4\delta_1\delta_2}{(1+\delta_1)(1+\delta_2)}\right)^{1/2},
\eeqa
where $\delta_l=1-2n_l$.

%----------------------------------
%\subsection{Electron- and hole-doped layers}
%----------------------------------
If the first plane is a hole-doped one without double occupancy and the 2nd plane is an electron-doped one without hole, i.e., $1-2n_2+d_2=0$, the interlayer GW factor is given by
\beqa
&&q_{\perp}=\left(\frac{4\delta_1d_2}{(1+\delta_1)(1+d_2)}\right)^{1/2},
\eeqa
where $1-n_2=n_2-d_2$.

\section{Fermi Surfaces in Four-Layered Systems}
\subsection{Multilayered \tj model}
To examine the FS's in four-layered cuprates, we adopt multilayered \tj model\cite{tohyama,mori02} composed of two IP's and two OP's as shown in Fig. 2. 
The Hamiltonian is given by
\beqa
&& H 
	=H_{\parallel}+H_{\perp}+H_{W},\label{htot}\\
&&H_{\parallel}
	=
	\sum_{l=1}^4 \Bigl(-t\sum_{\langle ij\rangle_{1st},\sigma}
		\hat{c}^{(l)\dag}_{j,\sigma}\hat{c}^{(l)}_{i,\sigma}
	-t'\sum_{\langle ij\rangle_{2nd},\sigma}
		\hat{c}^{(l)\dag}_{j,\sigma}\hat{c}^{(l)}_{i,\sigma}\nonumber\\
&&-t''\sum_{\langle ij\rangle_{3rd},\sigma}
		\hat{c}^{(l)\dag}_{j,\sigma}\hat{c}^{(l)}_{i,\sigma}
+   J\sum_{\langle ij\rangle_{1st}}
		\vec{S}^{(l)}_i \cdot \vec{S}^{(l)}_j\Bigr), \label{intra}\\
&&H_{\perp}
	=
	\sum_{l\neq m,k,\sigma} \epsilon_{\perp}(k)
	\left(\hat{c}^{(l)\dag}_{k,\sigma}\hat{c}^{(m)}_{k,\sigma} + H.c.\right),\label{inter}\\
&&H_{W}
	=
	\frac{W}{2}\!\!\!\!\sum_{l\in{\rm IP},m\in{\rm OP}}\!\!\!\!
	\left( N_{l} - N_{m} \right), \label{site}\\
&&N_l=\sum_i n^{(l)}_i,
\eeqa
where $\hat{c}^{(l)}_{i,\sigma}=c^{(l)}_{i,\sigma}(1-n^{(l)}_{i,-\sigma})\,$ is the annihilation operator of an electron in the $l$-th layer with spin $\sigma$ at site $i$ with the constraint of no double occupancy, and $n^{(l)}_i=n^{(l)}_{i,\uparrow}+n^{(l)}_{i,\downarrow}$ and $\vec{S}^{(l)}_{i}$ are the charge and the spin operators, respectively. 

In eq. (\ref{intra}), the summations $\langle ij\rangle_{1st}$, $\langle ij\rangle_{2nd}$ and $\langle ij\rangle_{3rd}$ run over first, second and third nearest-neighbor sites, respectively. 
The values of the parameters are as follows; $J=0.14$ eV, $t/J=2.5$, $t'/J=-0.85$ and $t''/J=0.575$.\cite{tohyama}

The inter-layer hopping in eq. (\ref{inter}) has the dispersion relation,\cite{chak_sci}
\beq
\epsilon_{\perp}(k)
	=
	-\frac{t_{\perp}}{4}
	\left(
	\cos(k_x)-\cos(k_y)
	\right)^2,\label{interhop}
\eeq
where $t_{\perp}$ is the amplitude without renormalization and is set to $t_{\perp}/J=1.0$\cite{chak_sci,okandersen,liechtenstein,feng,chuang}.

The chemical potential, $\mu$, is introduced to fix the total density as, $N=N_{\rm IP}+N_{\rm OP}$. The charge imbalance is measured by a difference of density in the IP and that in the OP, i.e., $\delta N \equiv N_{\rm IP}-N_{\rm OP}$. $\delta N$ is determined by a site-potential in eq. (\ref{site}), $W$, and the effective interlayer hopping, $t_{\rm eff} = q_{\perp}t_{\perp}$. 
In this paper, a value of $W$ is self-consistently determined by fixing $\delta N$. 

The order parameter given by 
\beq
\chi^{(l)}_{\tau}
	\equiv
	\frac{1}{N}\sum_i
	\langle 
	c^{(l)\dag}_{i,\uparrow  }c^{(l)}_{i+\tau,\uparrow}
	+c^{(l)\dag}_{i,\downarrow}c^{(l)}_{i+\tau,\downarrow}
	\rangle, \label{uRVB}
\eeq 
is introduced to the fourth term in eq. (\ref{intra}). 
The bond directions are indicated by $\tau=x, y$. 
The parameters, $\mu$, $W$, and $\chi^{(l)}_{\tau}$, are self-consistently determined by numerical calculations. 

We adopt the Gutzwiller approximation shown in \S2 to take the constraint of no double occupancy in the Hamiltonian (\ref{htot}). 
As a result, the electron creation- and annihilation operators are released from the constraint, and the parameters are changed as follows: 
\beqa
t &\to& q_t^{(l)} t, \;\;
t' \to q_t^{(l)} t', \;\;
t'' \to q_t^{(l)} t'', \\
J &\to& q_J^{(l)} J,\\
t_{\perp}&\to& t_{\perp 1}\equiv q_{\perp 1} t_{\perp}, \; t_{\perp 2}\equiv q_{\perp 2} t_{\perp},
\eeqa
with
\beqa
q_t^{(l)}&=& \frac{2\delta_l}{1+\delta_l},\\
q_J^{(l)}&=& \frac{4}{(1+\delta_l)^2},\\
q_{\perp 1} &=&\frac{2\delta_{\rm IP}}{1+\delta_{\rm IP}},\\
q_{\perp 2} &=&\left(\frac{4\delta_{\rm IP}\delta_{\rm OP}}{(1+\delta_{\rm IP})(1+\delta_{\rm OP})}\right)^{1/2}.
\eeqa 
If the IP becomes electron-doped, $\delta_{\rm IP}$ is replaced by $d_{\rm IP}$. 
The effective amplitudes of interlayer hoppings are denoted by $t_{\perp 1}\equiv q_{\perp 1} t_{\perp}$ between IP and OP, and $t_{\perp 2}\equiv q_{\perp 2} t_{\perp}$ between IP's.
%******
\begin{figure}[p]
\begin{center}
\includegraphics[width=7.5cm]{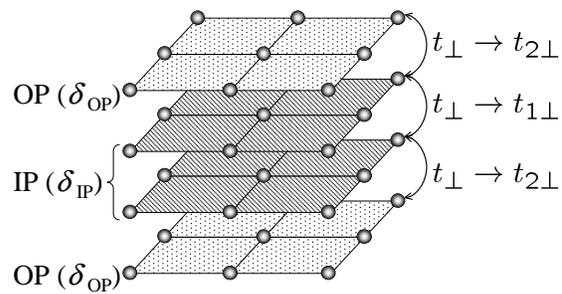}
%\vspace{10cm}
\caption{Schematic figure of four-layered system. The effective amplitude of interlayer hoppings are denoted by $t_{\perp 1}\equiv q_{\perp 1} t_{\perp}$ between IP and OP, and by $t_{\perp 2}\equiv q_{\perp 2} t_{\perp}$ between IP's. 
}
\label{fig2}
\end{center}
\end{figure}
%******
%---------------------------------- 
\subsection{Fermi surfaces}
%----------------------------------
%******
\begin{figure}[h]
\begin{center}
\includegraphics[width=6.5cm]{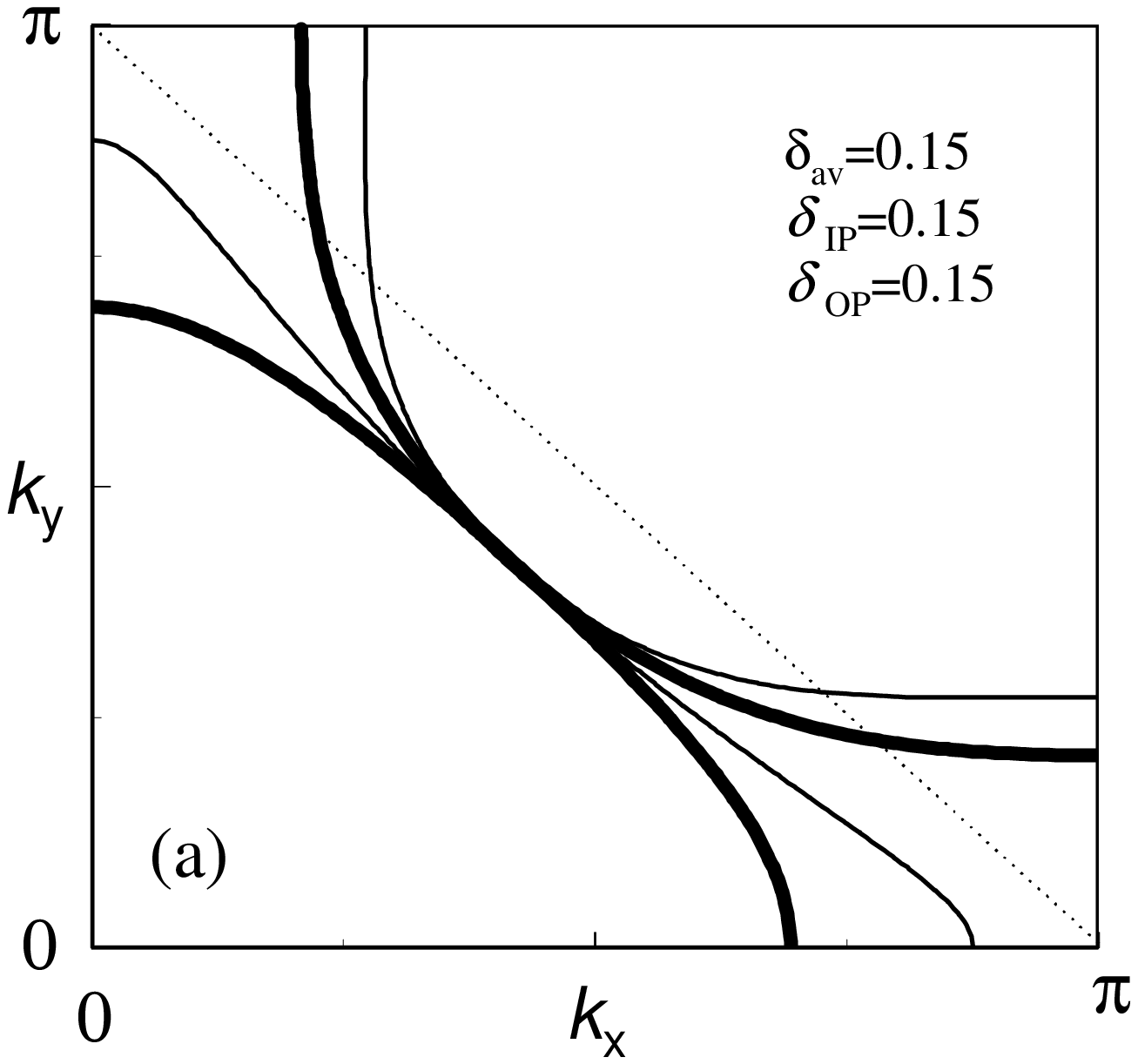}
\includegraphics[width=6.5cm]{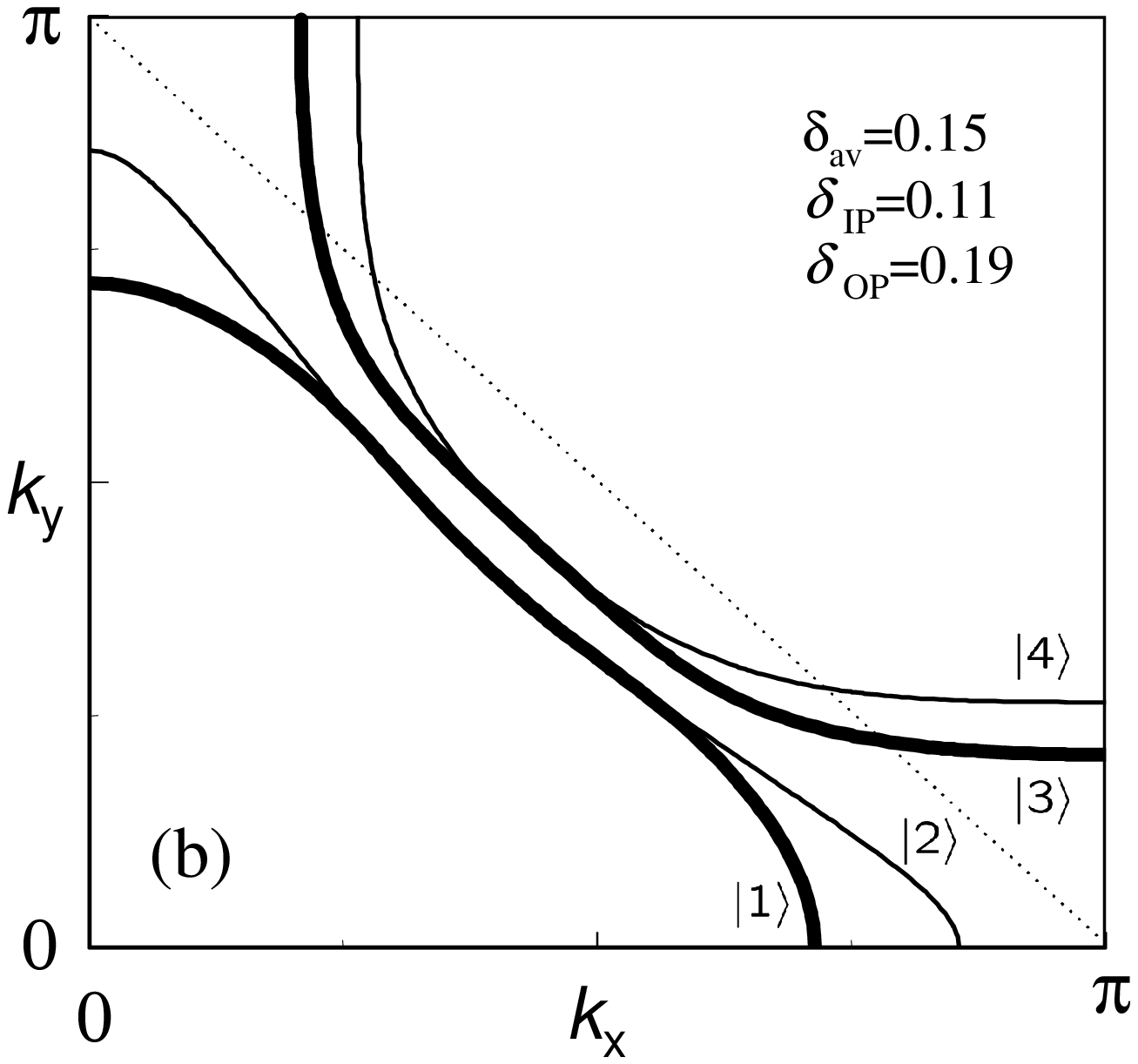}
\includegraphics[width=6.5cm]{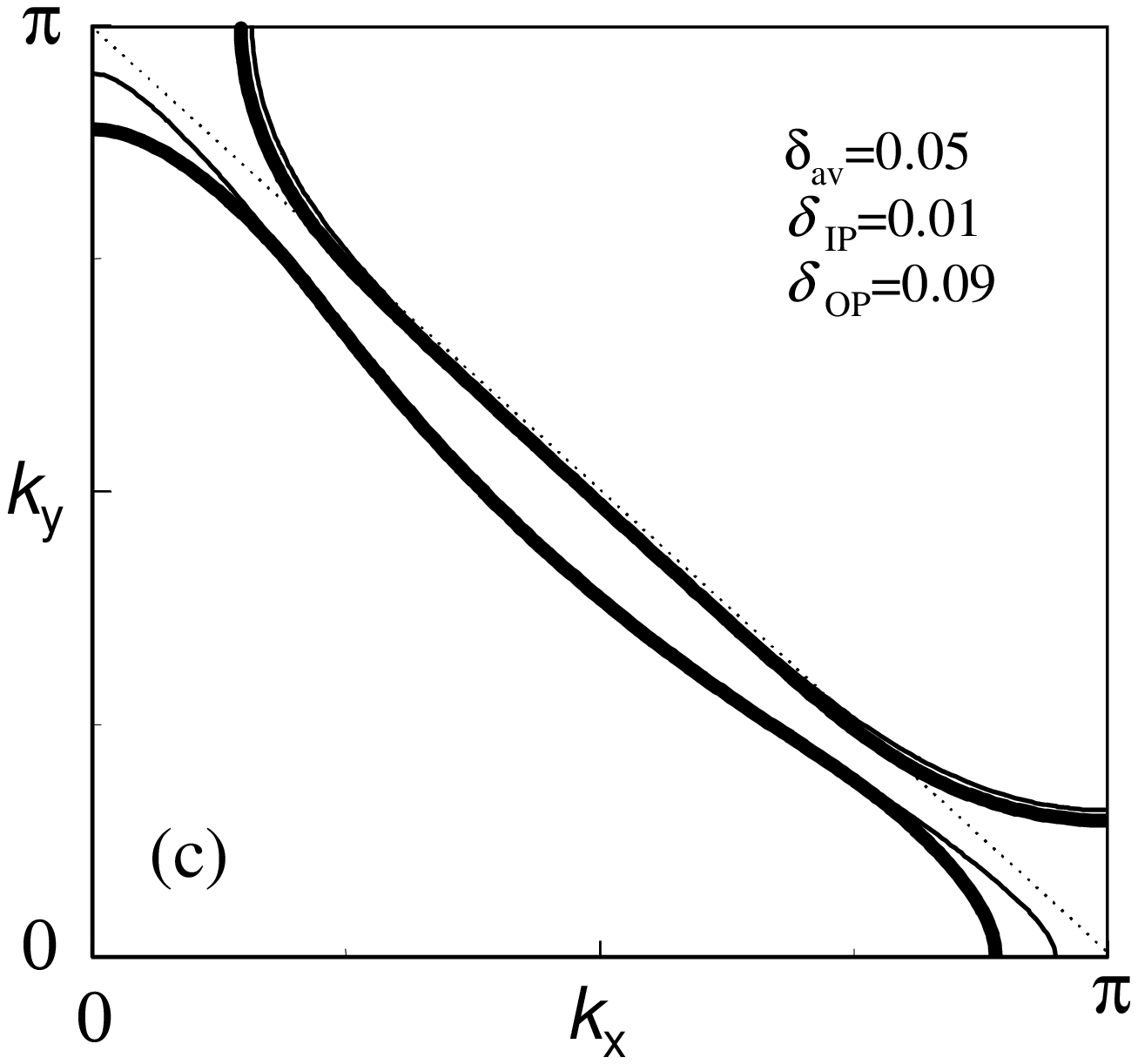}
\includegraphics[width=6.5cm]{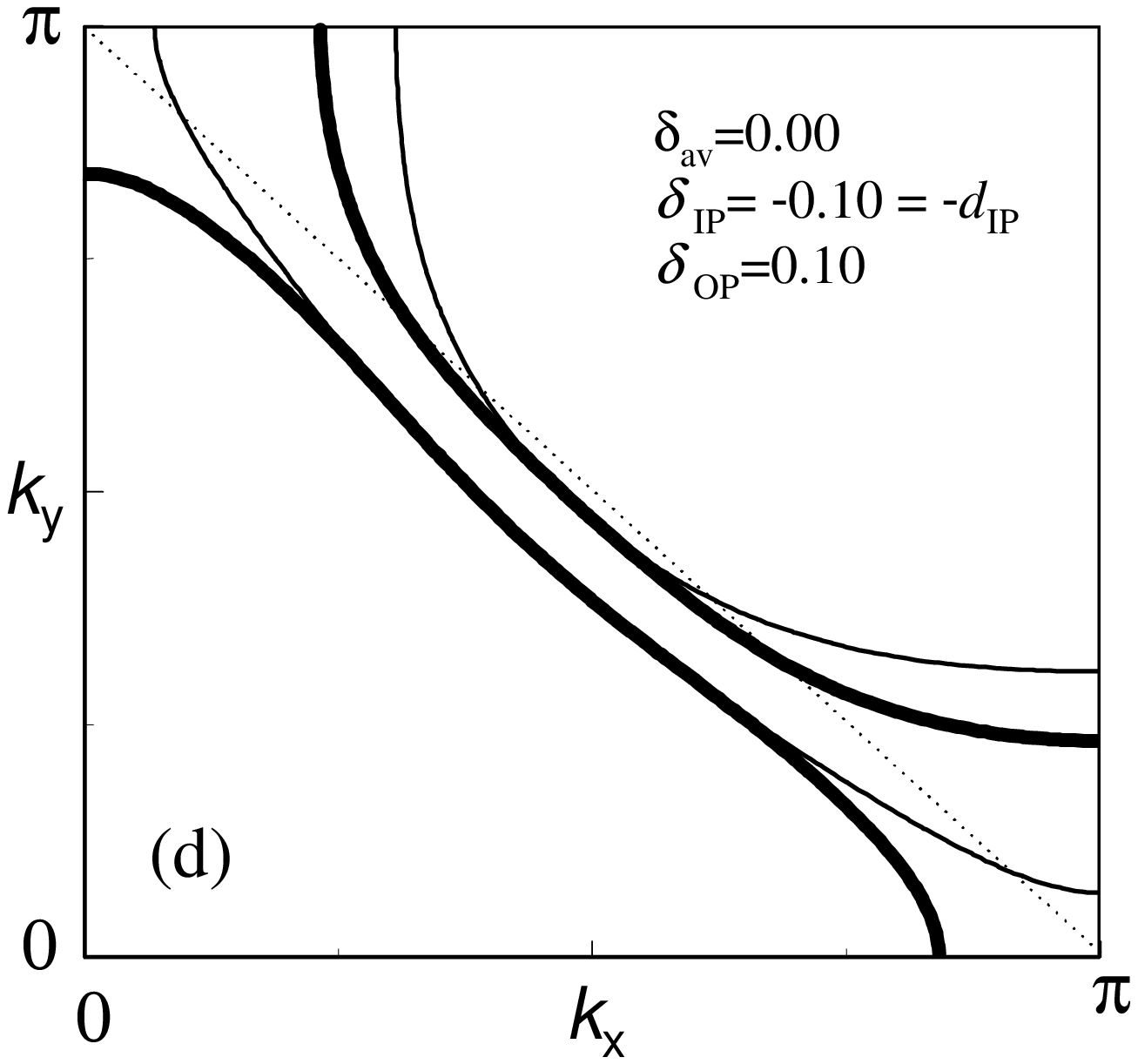}
\end{center}
\end{figure}
%******
\begin{figure}[h]
\begin{center}
%\includegraphics[width=6.5cm]{fig3a.eps}
%\includegraphics[width=6.5cm]{fig3b2.eps}
%\includegraphics[width=6.5cm]{fig3c.eps}
%\includegraphics[width=6.5cm]{fig3d2.eps}
%\vspace{10cm}
\caption{Fermi surfaces of four-layered systems with $J=0.14$ eV, $t/J=2.5$, $t'/J=-0.85$, $t''/J=0.575$, $t_{\perp}/J=1.0$ for 
(a) $\delta_{\rm av}$=0.15, $\delta_{\rm IP}$=0.15, $\delta_{\rm OP}$=0.15, 
(b) $\delta_{\rm av}$=0.15, $\delta_{\rm IP}$=0.11, $\delta_{\rm OP}$=0.19,  
(c) $\delta_{\rm av}$=0.05, $\delta_{\rm IP}$=0.01, $\delta_{\rm OP}$=0.09, and
(d) $\delta_{\rm av}$=0.00, $\delta_{\rm IP}$=-0.10, $\delta_{\rm OP}$=0.10.  
The averaged hole-density, $\delta_{\rm av}$, is defined as, $\delta_{\rm av}=(\delta_{\rm IP}+\delta_{\rm OP})/2$. 
Two thick (thin) lines correspond to the even (odd) symmetry band as regards the inversion. 
The doted line shows a half of Brillouin zone as a guide for eyes. 
The symbols, $|n\rangle$ (n=1-4), will be explained below. 
}
\label{fig3}
\end{center}
\end{figure}
%******

Fermi surfaces (FS's) of four-layered cuprates are shown in Fig. 3 for 
(a) $\delta_{\rm av}$=0.15, $\delta_{\rm IP}$=0.15, $\delta_{\rm OP}$=0.15, 
(b) $\delta_{\rm av}$=0.15, $\delta_{\rm IP}$=0.11, $\delta_{\rm OP}$=0.19,  
(c) $\delta_{\rm av}$=0.05, $\delta_{\rm IP}$=0.01, $\delta_{\rm OP}$=0.09, and
(d) $\delta_{\rm av}$=0.00, $\delta_{\rm IP}$=-0.10, $\delta_{\rm OP}$=0.10.
The averaged hole density, $\delta_{\rm av}$, is defined as $\delta_{\rm av}\equiv(\delta_{\rm IP}+\delta_{\rm OP})/2$. 

In the case of no charge imbalance given by $\delta_{\rm IP}$=0.15 and $\delta_{\rm OP}$=0.15, the FS does not split along the nodal direction as  shown in Fig. 3 (a), since $\epsilon_{\perp}(k)$ in eq. (\ref{interhop}) becomes zero in this direction. 
If one assumed a '$k$-independent' interlayer-hopping ($t_{\perp}^0$), the FS could split along the nodal direction as well. 
However,  by taking account of the experimental results about the bilayer cuprates, an amplitude of $t_{\perp}^0$ must be small compared to $t_{\perp}$\cite{feng, chuang}. 

On the other hand, in Fig. 3 (b), the FS along the nodal direction splits due to $W$ that induces the charge imbalance given by $\delta_{\rm IP}$=0.11 and $\delta_{\rm OP}$=0.19. 
The Fermi points in this direction are determined by 
$0=\epsilon_l (k) \pm W/2 -\mu$, where 
$\epsilon_l (k)\equiv -4t_l\cos k-4t'_l(\cos k)^2 -4t''_l\cos 2k -(3/2)J_l\chi_l\cos k$,
$t_l \equiv q_t^{(l)} t$, $t'_l \equiv q_t^{(l)} t'$, $t''_l \equiv q_t^{(l)} t''$, and $J_l\equiv q^{(l)}_J J$. 
The width of the splitting is roughly estimated as $\sum_l W/(8t_l+3J_l\chi_l)$ for $t'=t''=0$, $\mu=0$ and $W/t_l, W/(J_l\chi_l)\ll 1$. 
Without the charge imbalance, the FS does not split along the nodal direction, since $W\to 0$ and $\epsilon_{\perp}(k)$ in eq. (\ref{interhop}) becomes zero in this direction. 
Therefore, one can say that the FS splitting along the nodal direction stands for the charge imbalance. 

In both cases, Figs. 3 (a) and (b), the systems have enough hole density as $\delta_{\rm av}$=0.15, which splits the FS into four. 
Two of them look like a hole-like FS enclosing ($\pi$, $\pi$) point, while the other two enclose (0, 0) point like an electron-like case. 
Such two different types of FS originate from a fact that the dispersion relation is flat around the antinodal points, i.e., ($\pm \pi$, 0) and (0, $\pm \pi$). 
As a result, a tiny increase of electron density leads to a large shift of FS. 
Particularly, since an amplitude of  $\epsilon_{\perp}(k)$ is maximum at the antinodal points, the interlayer splitting is large and then results in the electron- and the hole-like FS's. 
Note that there is no electron-doped \cuo plane in Figs. 3 (a) and (b). 

By decreasing the averaged hole density as $\delta_{\rm av}$=0.05 and imposing the charge imbalance given by $\delta_{\rm IP}$=0.01 and $\delta_{\rm OP}$=0.09, the number of the FS's seems to approach to two as shown in Fig. 3 (c); one is hole-like FS and the other one is electron-like FS.  The splitting of the two hole-like (electron-like) FS's is determined by $t_{\perp 1}/J_{\rm IP}\chi_{\rm IP}$, which is proportional to $\delta_{\rm IP}$, and $t_{\perp 1}/t_{\perp 2}=\sqrt{\delta_{\rm IP}/\delta_{\rm OP}}$. 

So far, we have discussed the hole-doped cases given by $\delta_{\rm av}>0$. 
Assumed that all apical oxygens are ideally substituted by fluorine like Ba$_2$Ca$_3$Cu$_4$O$_8$F$_2$, the nominal valence of Cu becomes +2.
In such a case, we can expect the half-filled system as $\delta_{\rm av}$=0.0, but the system has the charge imbalance given by $\delta_{\rm IP}$=-$d_{\rm IP}$ and $\delta_{\rm OP}=d_{\rm IP}$. 
Note that the IP becomes electron doped. 
In Fig. 3 (d), we show FS's in such a case with $\delta_{\rm OP}$=$d_{\rm IP}=0.10$. 
The FS's are split into four as is the case of Fig. 3 (b). 

The splitting or the degeneracy of FS's is explained by a simple energy diagram shown in Fig. 4. 
%******
\begin{figure}[h]
\begin{center}
\includegraphics[width=8.6cm]{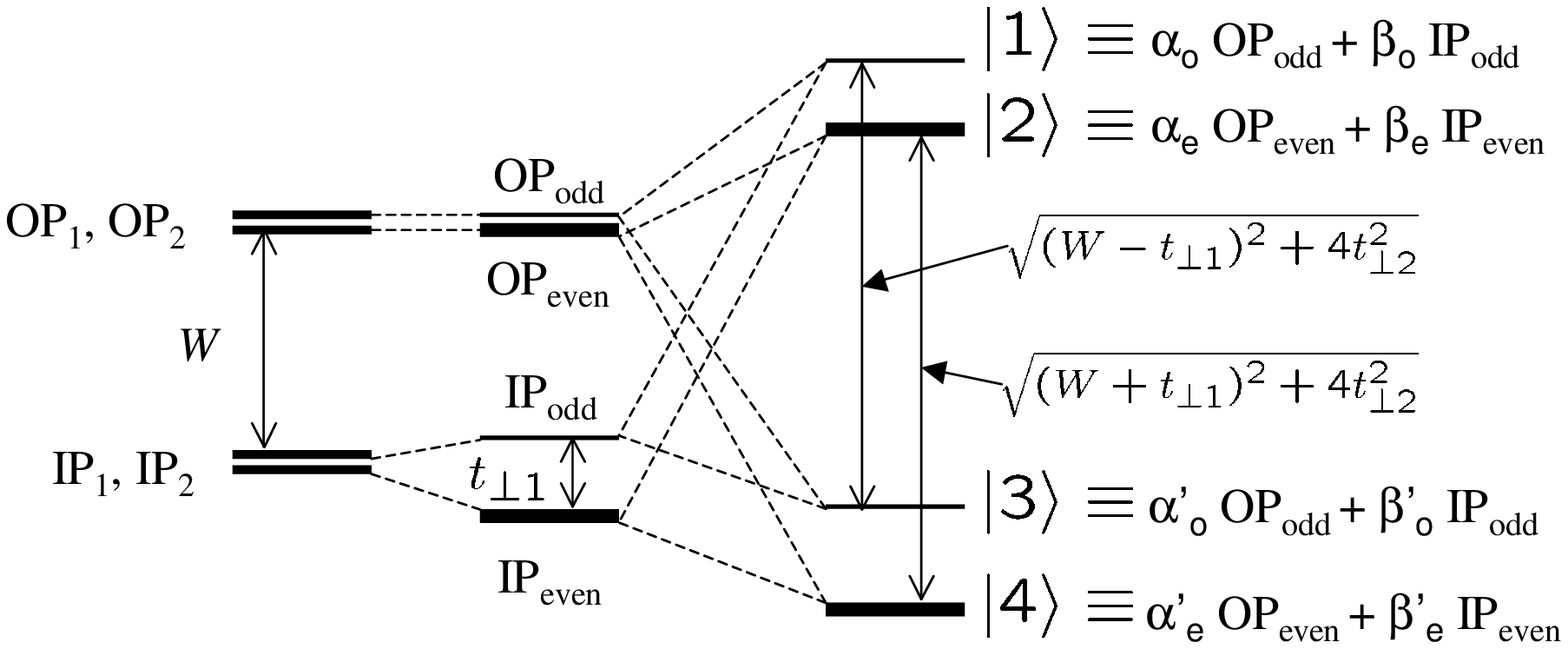}
%\vspace{10cm}
\caption{Energy scheme of four-layers composed of OP$_1$, OP$_2$, IP$_1$ and IP$_2$. The energy of OP$_1$ and OP$_2$ is separated from IP$_1$ and IP$_2$ by $W$. Considering the inversion symmetry, these planes are classified into even- and odd wave function as OP$_{\rm even}$, OP$_{\rm odd}$, IP$_{\rm even}$ and IP$_{\rm odd}$. Including $t_{\perp 1}$, the degenerated states of IP$_{\rm even}$ and IP$_{\rm odd}$ are separated. 
}
\label{fig4}
\end{center}
\end{figure}
%******
Without the interlayer hoppings, the degenerated OP's are located at $W$ in energy above the degenerated IP's. 
By considering the inversion symmetry, each pair of planes is classified into even- and odd wavefunctions given by OP$_{\rm even, odd}\equiv$ (OP$_1$$\pm$OP$_2$)/$\sqrt{2}$ and IP$_{\rm even, odd}\equiv$ (IP$_1$$\pm$IP$_2$)/$\sqrt{2}$. 
The IP$_{\rm even}$ and IP$_{\rm odd}$ are separated by $t_{\perp 1}$, which finally leads to the splitting between the hole-like (electron-like) FS's. 
The remaining degeneracy is lifted by $t_{\perp 2}$. 
Consequently, we obtain four separated FS's, $|1\rangle$, $|2\rangle$, $|3\rangle$, and $|4\rangle$, from  (0, 0) to ($\pi$, $\pi$) points as indicated in Fig. 3 (b). 
Each wavefunction is given by the linear combination as, 
$|1\rangle\equiv\alpha_{\rm o}$OP$_{\rm odd}$+$\beta_{\rm o}$IP$_{\rm odd}$, 
$|2\rangle\equiv\alpha_{\rm e}$OP$_{\rm even}$+$\beta_{\rm e}$IP$_{\rm even}$, 
$|3\rangle\equiv\alpha'_{\rm o}$OP$_{\rm odd}$+$\beta'_{\rm o}$IP$_{\rm odd}$, and 
$|4\rangle\equiv\alpha'_{\rm e}$OP$_{\rm even}$+$\beta'_{\rm e}$IP$_{\rm even}$. 
The splitting between the hole-like (electron-like) FS's in Figs. 3 (a)-(c) is identified with that between $|3\rangle$ and $|4\rangle$ ($|1\rangle$ and $|2\rangle$). 
Although the hole-like FS's are mainly composed of IP's wavefunction, the OP's wavefunction also contribute to the hole-like FS's. 
As a result, for example, a ratio of the area of FS $|4\rangle$ to the Brillouin zone is equivalent to the electron density given by $\alpha_{\rm o}'^2 n_{\rm OP}$+$\beta_{\rm o}'^2 n_{\rm IP}$, where $n_{\rm OP}$ and $n_{\rm IP}$ are the electron densities in the OP and the IP, respectively. 

%---------------------------------- 
\section{Summary and Discussion}
%----------------------------------
In this paper, we have examined the charge imbalance effects on Fermi surface (FS) splittings of multilayered cuprates by the multilayered \tj model in the  mean-field theory. 
Amplitude of interlayer hoppings is renormalized by the electron-electron interaction in the \cuo plane. 
In order to take such a correlation effect in the calculation, we have extended the Gutzwiller approximation to the interlayer hopping and have shown its Gutzwiller factor. 

It is found that the splitting of FS's along the nodal direction stands for the charge imbalance, which leads to the inhomogeneous interlayer hoppings. 
Considered no double occupancy in the \cuo plane in the four-layered system, the Gutzwiller factor, $q_{\perp}$, is proportional to the square root of the product of the doping rates in neighboring two planes, resulting in two different values of $q_{\perp}$, 
i.e., $q_{\perp,1}\propto \sqrt{\delta_{\rm IP}\delta_{\rm IP}}$ 
between inner planes (IP's), and 
$q_{\perp,2}\propto \sqrt{\delta_{\rm IP}\delta_{\rm OP}}$ 
between IP and outer plane (OP). 
With decreasing a hole density in the IP, four FS's in the four-layered system degenerate into two due to $q_{\perp,1}\to 0$. 

In this study, we have not considered the antiferromagnetic (AF) long- nor short-ranged order, which induces gapped behavior in the single-particle spectral function, $A(k,\omega)$, near the antinodal region\cite{tohyama94,tohyama04}. 
The gapped behavior appears in the FS observed by ARPES as an electron pocket\cite{yoshida,kmshen}. 
Considered the charge imbalance in the underdoped region, the IP's may have the AF short- or long-ranged order, and are connected to the paramagnetic- or the superconducting (SC) OP's by the interlayer hoppings. 
To compare our results with experimentally determined FS's, it may be necessary to take the AF order in the calculations but leave them as future issues. 

\section*{Acknowledgment}
We would like to thank H. Eisaki, A. Iyo, H. Mukuda, K. Tokiwa for useful discussions and showing us their experimental data. 
We would like to thank G. Seibold for his valuable suggestions and discussions. 
This work was supported by (a Grand-in-Aid for Scientific Research on Priority Areas and) the NAREGI Nanoscience Project from MEXT and CREST. 
One of authors (M. M.) acknowledges support by 21st Century COE program, Tohoku University,  Materials Research Center and Kakenhi Wakate (B) from MEXT.

\appendix
%-------------------------------------------------------------------
\section{Gutzwiller Approximation with Inter-Site Correlation}
%-------------------------------------------------------------------
The derivation of eq. (\ref{qperp}) is summarized in this Appendix. 
We apply the method developed by Ogawa and Kanda (ref. 44) to include the intersite correlation into $q_{\perp}$. 

First, we consider two planes with $L$ sites per plane. 
The number of up- and down spins in $l$-th plane are denoted by $M_l$ and $N_l$, respectively.  Given constants, $L$, $M_l$ and $N_l$, satisfy the following relations as 
\beqa
L &=& D_l +A_l +B_l +E_l,\\
M_l&=& A_l +D_l,\\
N_l&=& B_l+D_l,
\eeqa
where $D_l$ and $E_l$ are the number of the double-occupied and the vacant sites in the $l$-th plane, respectively. 
$A_l$ ($B_l$) means the number of single-occupied sites in the $l$-th plane with up (down) spin.  
Next, one divides the system into pairs of two sites, in which one belongs to 1st plane and the other one does to 2nd plane. 
Each pair has various configurations, e.g. one is up-spin site and the other one is doubly occupied site. 
The number of pairs with $s_1$-site in the 1st plane and $s_2$-site in the 2nd one are denoted by $y_{s_1s_2}~(s_l = S_l/L, S_l = D_l, A_l, B_l, E_l)$. 
These quantities, $y_{s_1 s_2}$, satisfy the following relations, 
\beqa
p_{1}&=&y_{d_1d_2} + y_{d_1a_2} + y_{a_1d_2} + y_{a_1a_2},\\
p_{2}&=&y_{d_1b_2} + y_{d_1e_2} + y_{a_1b_2} + y_{a_1e_2},\\
p_{3}&=&y_{b_1d_2} + y_{e_1d_2} + y_{b_1a_2} + y_{e_1a_2},\\
p_{4}&=&y_{b_1b_2} + y_{b_1e_2} + y_{e_1b_2} + y_{e_1e_2},\\
q_{1}&=& y_{d_1d_2} + y_{d_1b_2} + y_{b_1d_2} + y_{b_1b_2},\\
q_{2}&=& y_{d_1a_2} + y_{d_1e_2} + y_{b_1a_2} + y_{b_1e_2},\\
q_{3}&=& y_{a_1d_2} + y_{e_1d_2} + y_{a_1b_2} + y_{e_1b_2},\\
q_{4}&=& y_{a_1a_2} + y_{a_1e_2} + y_{e_1a_2} + y_{e_1e_2},\\
s_1 &=& \sum_{s_2=d_2, a_2, b_2, e_2} y_{s_1 s_2},\\
s_2 &=& \sum_{s_1=d_1, a_1, b_1, e_1} y_{s_1 s_2}, 
\eeqa
where $p_i=P_i/L$ and $q_i=Q_i/L$. 
The parameters, $p_i$, $q_i$, and $y_{s_1 s_2}$, are determined by optimizing the wavefunction that satisfies the following relations, 
\beqa
\langle\Psi|\Psi\rangle
	&=& \sum_{D_l,{\cal M}_i {\cal N}_i} g_1^{2D_1}g_2^{2D_2}
				\det {\cal U}\det{\cal V},\label{norm}\\
\det {\cal U}
	&=&
		\prod_{l=1,2} \omega_{M_l}^{M_l}\omega_{\bar{M}_l}^{L-M_l}
		\prod_{j=1}^4 \alpha_j^{P_j},\\
\det {\cal V}
	&=&
		\prod_{l=1,2}	\omega_{N_l}^{N_l}\omega_{\bar{N}_l}^{L-N_l}
		\prod_{j=1}^4 \beta_j^{Q_j},\\
\alpha_{1}
	&=& 1-\frac{|u|^2}{m_1m_2},\\
\beta_{1}
	&=& 1-\frac{|v|^2}{m_1m_2}\\
\alpha_{2}
	&=& 1+\frac{|u|^2}{m_1(1-m_2)},\\
\beta_{2}
	&=& 1+\frac{|v|^2}{m_1(1-m_2)}\\
\alpha_{3}
	&=& 1+\frac{|u|^2}{(1-m_1)m_2},\\
\beta_{3}
	&=& 1+\frac{|v|^2}{(1-m_1)m_2},\\
\alpha_{4}
	&=& 1-\frac{|u|^2}{(1-m_1)(1-m_2)},\\
\beta_{4}
	&=& 1-\frac{|v|^2}{(1-m_1)(1-m_2)},\\
u &=& \langle c^{(l)\dag}_{i\ua}c^{(m)}_{j\ua}\rangle_0
	 = \langle c^{(l)\dag}_{i\ua}c^{(m)}_{j\ua}\rangle_0,\\
v &=& \langle c^{(l)\dag}_{i\da}c^{(m)}_{j\da}\rangle_0
	 = \langle c^{(l)\dag}_{i\da}c^{(m)}_{j\da}\rangle_0.
\eeqa
The number of configurations that equally contribute to the summation in eq. (\ref{norm}) is given by
\beqa
\Gamma
	&=&
	\frac{L!}{\prod_{s_1,s_2} Y_{s_1s_2}!}.
\eeqa
In the limit of infinitely large system, the summation is replaced by dominant terms, and eq. (\ref{norm}) is approximated as 
\beq
\langle\Psi|\Psi\rangle
	\sim g_1^{2D^{(0)}_1} g_2^{2D^{(0)}_2}
				\frac{L!}{\prod_{s_1,s_2} Y^{(0)}_{s_1s_2}!}
				\det {\cal U}\det{\cal V}, 
\eeq
where $``(0)"$ indicate the dominant term. 
Below, we will abbreviate $``(0)"$ to simplify the notation. 
By optimizing the following function, 
\beqa
F(\{Y,P,Q\})
		&=&\ln\left [
				g_1^{2D_1} g_2^{2D_2}
				\frac{L!}{\prod_{s_1,s_2} Y_{s_1s_2}!}
				\det {\cal U}\det{\cal V}\right]\nonumber\\ 
		&+&\sum_{s_1=d_1,a_1,b_1,e_1} \ln \Lambda_{s_1}(\sum_{s_2} Y_{s_1s_2} - S_1),\nonumber\\
		&+&\sum_{s_2=d_2,a_2,b_2,e_2} \ln \Lambda_{s_2}(\sum_{s_1} Y_{s_1s_2} - S_2),\nonumber\\
		&+&\sum_{s_1,s_2,i} \ln \alpha'_{s_1,s_2}(\sum Y_{s_1,s_2} - P_i)\nonumber\\
		&+&\sum_{s_1,s_2,i} \ln \beta'_{s_1,s_2}(\sum Y_{s_1,s_2} - Q_i),\label{opt}
\eeqa
one can obtain the dominant terms in a factorized form as 
\beqa
Y_{s_1s_2} &=& \Lambda_{s_1} \Lambda_{s_2} \alpha_{s_1s_2} \beta_{s_1s_2}.
\eeqa
The parameter $\Lambda$'s introduced in eq. (\ref{opt}) as the Lagrange multiplier are determined by solving the simultaneous equation as, 
\beq
\begin{split}
D_1 &=
	 \Lambda_{d_1}\Lambda_{d_2}\alpha_1\beta_1
	+\Lambda_{d_1}\Lambda_{a_2}\alpha_1\beta_2
	+\Lambda_{d_1}\Lambda_{b_2}\alpha_2\beta_1\\
	&+\Lambda_{d_1}\Lambda_{e_2}\alpha_2\beta_2,\\
\end{split}
\eeq
\beq
\begin{split}
A_1 &= 
	 \Lambda_{a_1}\Lambda_{d_2}\alpha_1\beta_3
	+\Lambda_{a_1}\Lambda_{a_2}\alpha_1\beta_4
	+\Lambda_{a_1}\Lambda_{b_2}\alpha_2\beta_3\\
	&+\Lambda_{a_1}\Lambda_{e_2}\alpha_2\beta_4,\\
\end{split}
\eeq
\beq
\begin{split}
B_1 &= 
	 \Lambda_{b_1}\Lambda_{d_2}\alpha_3\beta_1
	+\Lambda_{b_1}\Lambda_{a_2}\alpha_3\beta_2
	+\Lambda_{b_1}\Lambda_{b_2}\alpha_4\beta_1\\
	&+\Lambda_{b_1}\Lambda_{e_2}\alpha_4\beta_2,\\
\end{split}
\eeq
\beq
\begin{split}
E_1 &=
	 \Lambda_{e_1}\Lambda_{d_2}\alpha_3\beta_3
	+\Lambda_{e_1}\Lambda_{a_2}\alpha_3\beta_4
	+\Lambda_{e_1}\Lambda_{b_2}\alpha_4\beta_3\\
	&+\Lambda_{e_1}\Lambda_{e_2}\alpha_4\beta_4,\\
\end{split}
\eeq
\beq
\begin{split}
D_2 &= 
	 \Lambda_{d_1}\Lambda_{d_2}\alpha_1\beta_1
	+\Lambda_{a_1}\Lambda_{d_2}\alpha_1\beta_3
	+\Lambda_{b_1}\Lambda_{d_2}\alpha_3\beta_1\\
	&+\Lambda_{e_1}\Lambda_{d_2}\alpha_3\beta_3,\\
\end{split}
\eeq
\beq
\begin{split}
A_2 &=
	 \Lambda_{d_1}\Lambda_{a_2}\alpha_1\beta_2
	+\Lambda_{a_1}\Lambda_{a_2}\alpha_1\beta_4
	+\Lambda_{b_1}\Lambda_{a_2}\alpha_3\beta_2\\
	&+\Lambda_{e_1}\Lambda_{a_2}\alpha_3\beta_4,\\
\end{split}
\eeq
\beq
\begin{split}
B_2 &= 
	 \Lambda_{d_1}\Lambda_{b_2}\alpha_2\beta_1
	+\Lambda_{a_1}\Lambda_{b_2}\alpha_2\beta_3
	+\Lambda_{b_1}\Lambda_{b_2}\alpha_4\beta_1\\
	&+\Lambda_{e_1}\Lambda_{b_2}\alpha_4\beta_3,\\
\end{split}
\eeq
\beq
\begin{split}
E_2 &= 
	 \Lambda_{d_1}\Lambda_{e_2}\alpha_2\beta_2
	+\Lambda_{a_1}\Lambda_{e_2}\alpha_2\beta_4
	+\Lambda_{b_1}\Lambda_{e_2}\alpha_4\beta_2\\
	&+\Lambda_{e_1}\Lambda_{e_2}\alpha_4\beta_4. 
\end{split}
\eeq
The optimizations about $D_l$ as $\partial F/\partial D_l =0$ lead to
\beqa
g_l^2 = \frac{\Lambda_{D_l}\Lambda_{E_l}}{\Lambda_{A_l}\Lambda_{B_l}}. 
\eeqa
Although there are several ways to get the Gutzwiller factor of hopping integral, we follow the way of Vollhardt (ref. 45). 
Thereby, the amplitude of interlayer hoppings are estimated as,
\beq
\begin{split}
&\langle\Psi| c^{(2)\dag}_{i\sigma} c^{(1)}_{j\sigma} |\Psi\rangle\\
	&=
	\frac{y_{a_1e_2} + g_2 y_{a_1b_2} + g_1y_{d_1e_2}  + g_1g_2 y_{d_1b_2} }
	       {\omega_{M_1}(1-\omega_{M_2})}
	       \langle\psi_0| c^{(2)\dag}_{i\sigma} c^{(1)}_{j\sigma} |\psi_0\rangle,
\end{split}
\eeq
\beq
\begin{split}
&\langle\Psi| c^{(1)\dag}_{i\sigma} c^{(2)}_{j\sigma} |\Psi\rangle\\
&=
	\frac{y_{e_1a_2} + g_1 y_{b_1a_2} + g_2 y_{e_1d_2} + g_2g_1 y_{b_1e_2} }
	       {\omega_{M_2}(1-\omega_{M_1})}
	       \langle\psi_0| c^{(1)\dag}_{i\sigma} c^{(2)}_{j\sigma} |\psi_0\rangle.
\end{split}
\eeq
Finally, we find that the Gutzwiller factor of the interlayer hopping is given by
\beq
q_{\perp}
	=
		\left(\frac{\langle\Psi| c^{(2)\dag}_{i\sigma} c^{(1)}_{j\sigma}|\Psi\rangle}
			  {\langle\psi_0| c^{(2)\dag}_{i\sigma} c^{(1)}_{j\sigma} |\psi_0\rangle}
		       \frac{\langle\Psi| c^{(1)\dag}_{i\sigma} c^{(2)}_{j\sigma}  |\Psi\rangle}
		       {\langle\psi_0| c^{(1)\dag}_{i\sigma} c^{(2)}_{j\sigma} |\psi_0\rangle}
		\right)^{1/2}.\label{qinter}
\eeq
In the case of $U\rightarrow \infty$,
\beq
q_{\perp}
	=
	\left(\frac{y_{a_1e_2} y_{e_1a_2}}{\omega_{M_1}(1-\omega_{M_2})\omega_{M_2}(1-\omega_{M_1})}
	\right)^{1/2}.\label{qinter2} 
\eeq
We have numerically solved the simultaneous equation (A.31)-(A.38) as a function of $u$ and $v$. 
It is found that the intersite correlations are negligible for small values of $u$ and $v$ about less than 0.2. 
In the multilayered \tj model introduced in \S3, $u$ and $v$ are smaller than 0.1, and then the site-approximation can be justified. 
Equation (\ref{qperp}) is obtained in the site approximation of eq. (\ref{qinter2}). 
In this case, $u=v=0$ and the following relations are obtained as,
\beqa
g_l^2
	&\rightarrow& \frac{d_l(1-2n_l+d_l)}{(n_l-d_l)^2}, \\
Y_{s_1s_2} &=& \Lambda_{s_1} \Lambda_{s_2} = L~y_{s_1s_2} = \frac{S_1S_2}{L}. 
\eeqa

%The number of equal contributions is approximately same as
%\beq
%\Gamma = \frac{L!}{\prod_{S_1} S_1!} \frac{L!}{\prod_{S_2} S_2!}.
%\eeq

\end{document}